\documentclass[aps,prl,twocolumn,showpacs]{revtex4-1}
\usepackage{graphicx,color,epstopdf,amsmath}

\begin{document}
\title{Baryonic $Z'$ Explanation for the CDF $Wjj$ Excess}
\author{ Kingman Cheung$^{1,2}$, Jeonghyeon Song$^{1}$}
\affiliation{
$^1$Division of Quantum Phases \& Devices, School of Physics, 
Konkuk University, Seoul 143-701, Republic of Korea \\
$^2$Department of Physics, National Tsing Hua University, 
Hsinchu 300, Taiwan
}

\renewcommand{\thefootnote}{\arabic{footnote}}
\date{\today}

\begin{abstract}
  The latest CDF anomaly, the excess of dijet events in the
  invariant-mass window $120-160$ GeV in associated production with a
  $W$ boson, can be explained by a baryonic $Z'$ model in which the
  $Z'$ boson has negligible couplings to leptons.  Although this $Z'$
  model is hardly subject to the Drell-Yan constraint from Tevatron,
  it is constrained by the dijet data from UA2 ($\sqrt{s} = 630$ GeV),
  and the precision measurements at LEP through the mixing with the SM
  $Z$ boson.  We show that under these constraints this model can
  still explain the excess in the $M_{jj} \sim 120 - 160$ GeV window,
  as well as the claimed cross section $\sigma(W Z') \sim 4$ pb.
  Implications at the Tevatron would be the associated production of
  $\gamma Z',\; Z Z'$, and $Z'Z'$ with the $Z' \to jj$. We show that
  with tightened jet cuts and improved systematic
  uncertainties both $\gamma Z' \to \gamma jj$ and $Z Z' \to \ell^+
  \ell^- j j$ channels could be useful to probe this model at the
  Tevatron.
\end{abstract}

\maketitle

{\it Introduction.}--
The year 2011 is perhaps the last year of running for the Tevatron,
which is subject to a severe budget cut. Hopefully, the recent surprises
\cite{cdf-top,cdf-wjj} from the Tevatron can reverse its fate. 
The latest surprise is an excess in the invariant-mass
window $120 - 160$ GeV in the dijet system of the associated production
of a $W $ boson with 2 jets \cite{cdf-wjj}.  
We shall denote it by $Wjj$ production.
The excess in the window $M_{jj} \sim 120 - 160$ GeV appears to be
a resonance, but the current resolution \cite{cdf-wjj}
cannot tell whether it is a narrow resonance.
From the distribution we can naively see that the width
of the resonance appears to be slightly wider than the SM $Z$ boson.

In this Letter, we propose a baryonic $Z'$ model to explain the anomaly. 
The reason for being baryonic is that even
if this $Z'$ has a small leptonic branching ratio,
even $O(1)\%$, it would suffer from strong constraints of the Tevatron
$Z'$ search in the dilepton mode \cite{cdf-z'}.
The baryonic $Z'$ model was proposed by Barger, Cheung, and Langacker 
in 1996 \cite{bcl} 
in light of the $R_{b}/R_{c}$ crisis of the LEP precision measurements
at that time \cite{lep}: 
$R_b = \Gamma(Z\to \bar bb) / \Gamma(Z\to \rm hadrons)$
deviated by $3.7\sigma$ while $R_c$ deviated by $-2.4\sigma$ from the
SM.  Through some adjustment of the mixing angle and vector and axial-vector
couplings the $R_b/R_c$ crisis can be solved (the most current data do
not show any more of the problem \cite{pdg}). Such a $Z'$ interpretation
at that time had suggested strong implications at the Tevatron \cite{bcl}
via $s$-channel $Z'$ production and the pair production
processes $(W,Z,\gamma)Z'$ with $Z'\to jj$ (in particular $b\bar b$),
with invariant mass $M_{jj}$ peaked at $M_{Z'}$.
The $s$-channel $Z'$ production is buried under the QCD background, but
the associated production with a $W$ boson has a good chance to appear.
The current CDF anomaly \cite{cdf-wjj} may be of this origin.

Additional $Z'$ bosons can appear in many extensions of the SM 
with extra $U(1)$'s \cite{paul}.
The most famous example is $E_6$, in which there are a number of extra 
neutral gauge bosons.  A baryonic $Z'$ can arise from a gauge
symmetry generated by the baryon number $U(1)_B$ as an interesting
possibility \cite{carone}, since this avoids potential 
problems associated with
the breaking of global baryon number by quantum gravity effects (e.g., an
unacceptable proton decay rate in supersymmetric theories). 
Another possibility is kinetic mixing of the two
$U(1)$'s \cite{babu} to suppress the leptonic couplings.
Here we assume that the model can be embedded in an anomaly-free theory. 

In this work, we use the baryonic $Z'$ model 
with $M_{Z'} \sim 145$
GeV to explain the excess of events in the invariant-mass window
of $M_{jj} \sim 120-160$ GeV in $Wjj$ production.  
The $Z'$ boson has negligible couplings to leptons, 
and so is not affected by
 the dilepton $Z'$ constraints.
However, it is constrained by
the dijet searches at hadronic colliders.
We found that all the dijet searches by CDF \cite{cdf-dijet} focused
on the mass region $M_{jj} > 200 $ GeV, and so the $Z'$ with $M_{Z'}
\sim 145$ GeV is not subject to these searches.  On the other hand, 
some old data from UA2 ($\sqrt{s} = 630$ GeV) \cite{ua2}
had better measurements in $M_{jj} = 100 - 200 $ GeV. We use 
the constraint on the coupling of the $Z'$ obtained in Ref.~\cite{bcl}.
Furthermore, the precision measurements at LEP also constrained 
the mixing with the SM $Z$ boson to be small $\le 10^{-3}$.  
We show that under these 
constraints this model can still explain the excess 
in the $M_{jj} \sim
120 - 160$ GeV window, as well as the claimed cross section 
$\sigma(W Z') \sim 4$ pb.  This is the main result of this work.
Further implications at the Tevatron would be the associated 
production of $\gamma Z',\; Z Z'$, and $Z'Z'$
with the $Z' \to jj$.
We show that it is hard to see the excess in
both $\gamma Z' \to \gamma jj$ and $Z Z' \to \ell^+ \ell^- jj$ channels
under the current level of systematic uncertainties and jet cuts.
However, with tightened jet cuts and improved systematic uncertainties
it could be promising to test the model in these two channels.

{\it The interactions.}--
Following Ref.~\cite{lang-luo}, the Lagrangian describing the
neutral current gauge interactions of the
standard electroweak $SU(2)\times U(1)$ and extra $U(1)$'s is given by
\begin{equation}
- {\cal L}_{\rm NC} = e J_{\rm em}^\mu A_\mu + \sum_{\alpha=1}^{n}
g_\alpha J^\mu_\alpha Z^0_{\alpha \mu}\;,
\end{equation}
where $Z^0_1$ is the SM $Z$ boson and $Z^0_\alpha$ with $\alpha\ge 2$ are the
extra $Z$ bosons in the weak-eigenstate basis.
For the present work we only consider one extra $Z_2^0$ 
mixing with the SM $Z^0_1$ boson. 
The coupling constant $g_1$ is the SM coupling $g/\cos\theta_{\rm w}$. 
For grand unified theories (GUT) $g_2$ is related to $g_1$ by
\begin{equation}
\frac{g_2}{g_1} = \left(\frac{5}{3}\, x_{\rm w} \lambda\right)^{1/2} \simeq
0.62\lambda^{1/2} \,,
\label{eq:g2/g1}
\end{equation}
where $x_{\rm w}=\sin^2\theta_{\rm w}$ and $\theta_{\rm w}$ is the weak mixing
angle.
The factor $\lambda$ depends on the symmetry breaking pattern and the fermion
sector of the theory, which is usually of order unity.

Since we only consider the mixing of $Z_1^0$ and $Z_2^0$ we can rewrite
the Lagrangian with only the $Z^0_1$ and $Z^0_2$ interactions
\begin{eqnarray}
-{\cal L}_{Z^0_1 Z^0_2} &=& g_1 Z^0_{1\mu} \left[ \frac{1}{2} \sum_i
 \bar \psi_i \gamma^\mu (g_v^{i(1)} - g_a^{i(1)} \gamma^5 ) \psi_i \right] 
 \nonumber \\
&+&
 g_2 Z^0_{2\mu} \left[ \frac{1}{2} \sum_i
 \bar \psi_i \gamma^\mu (g_v^{i(2)} - g_a^{i(2)} \gamma^5 ) \psi_i \right]\;,
\end{eqnarray}
where for both quarks and leptons
$
g_v^{i(1)} = T_{3L}^i - 2 x_{\rm w} Q_i \,, \qquad
g_a^{i(1)} = T_{3L}^i\,,
$
and we consider the case in which $Z^0_2$ couples only to quarks,
\begin{equation}
g_v^{q(2)} = \epsilon_V \,,\qquad
g_a^{q(2)} = \epsilon_A \,,\qquad
g_v^{\ell(2)} = g_a^{\ell(2)} = 0\;.
\end{equation}
Here $T_{3L}^i$ and $Q_i$ are, respectively, the third component of the weak
isospin and the electric charge of the fermion $i$.
The parameters $\epsilon_V$ and $\epsilon_A$ are the vector 
and axial-vector couplings of $Z^0_2$.
Without loss of generality we choose $\epsilon_V=\sin\gamma$ and  
$\epsilon_A=\cos\gamma$ such that $(\epsilon_V^2 + \epsilon_A^2) $ is
normalized to unity.
The mixing of the weak eigenstates $Z^0_1$ and $Z^0_2$ to form  mass 
eigenstates $Z$ and $Z'$  are parametrized
by a mixing angle $\theta$:
\begin{equation}
\label{mixing}
\left ( \begin{array}{c} Z \\
                         Z'
        \end{array} \right ) = \left( \begin{array}{rr}
                                  \cos\theta & \sin\theta \\
                                 -\sin\theta & \cos\theta
                                      \end{array} \right ) \;
    \left( \begin{array}{c} Z^0_1 \\
                            Z^0_2
            \end{array} \right ) \;.
\end{equation}
The mass of $Z$ is $M_{Z}=91.19$~GeV.

After substituting  the interactions of the mass eigenstates $Z$ and  
$Z'$ with fermions are 
\begin{eqnarray}
-{\cal L}_{Z Z' } &=& \sum_i \frac{g_1}{2} \biggl [
  Z_{\mu} \bar \psi_i \gamma^\mu (v_s^i - a_s^i \gamma^5 ) \psi_i  \nonumber \\
 && +
  Z'_{\mu} \bar \psi_i \gamma^\mu (v_n^i - a_n^i \gamma^5 ) \psi_i  \biggr ]\,,
\label{rule}
\end{eqnarray}
where
\begin{eqnarray}
v_s^i = g_v^{i(1)} + \frac{g_2}{g_1} \, \theta \, g_v^{i(2)} \,, &\qquad&
a_s^i = g_a^{i(1)} + \frac{g_2}{g_1} \, \theta \, g_a^{i(2)} \,, \\
v_n^i = \frac{g_2}{g_1}\, g_v^{i(2)} - \theta \, g_v^{i(1)}\,, &\qquad&
a_n^i = \frac{g_2}{g_1}\, g_a^{i(2)} - \theta \, g_a^{i(1)}\,.
\end{eqnarray}
Here we have used the valid approximation $\cos\theta\approx 1$ 
and $\sin\theta \approx \theta$.  In the following, we ignore the mixing 
($\theta =0$) such that the precision measurements for the SM $Z$ boson are not
affected, unless stated otherwise. 
We also take the democratic 
choice of equal couplings of $Z'$ to up-type and down-type quarks. This is in 
accord with the CDF observation that there is no preference for $b$ quarks
in the dijet window $M_{jj}= 120-160$ GeV \cite{cdf-wjj}.

{\it $s$-Channel $Z'$ production.}--
The decay width of
$Z' \to f \bar f$ is given by
\begin{eqnarray}
\Gamma (Z' \to f\bar f ) &=& \frac{G_F M_{Z}^2 }{6\pi \sqrt{2} }
N_c  C(M_{Z'}^2) M_{Z'} \sqrt{ 1 - 4x} \nonumber \\
&\times& \left[
v_{n}^{f2} (1+2x) +  a_{n}^{f2} (1-4x) \right] \,,
\label{width}
\end{eqnarray}
where $G_F$ is the Fermi coupling constant,
$C(M_{Z'}^2) = 1+\alpha_s/\pi + 1.409 (\alpha_s/\pi)^2 -12.77 (\alpha_s/\pi)^3$, 
$\alpha_s = \alpha_s (M_{Z'})$ is the strong coupling at the scale $M_{Z'}$,
$x=m_f^2/M_{Z'}^2$, 
and $N_c=3$ or 1 if $f$ is a quark or a lepton,
respectively.
The $Z'$ width is proportional to $\lambda$, which sets the strength of
the $Z'$ coupling. For $\lambda=1$ the total $Z'$ width is
$
\Gamma_{Z'} / M_{Z'} = 0.022 \quad {\rm for}\ M_{Z'} < 2m_t
$.
The width would be increased somewhat if there are open channels for decay
into the top quark, superpartners, and other exotic particles. 
Essentially, it is a narrow resonance.

The $Z'$ boson can be directly produced at a hadron collider via the $q\bar
q\to Z'$ subprocess, for which the cross section in the narrow $Z'$ width
approximation is \cite{collider}
\begin{equation}
\hat\sigma(q\bar q\to Z') = K \frac{2\pi}{3} \frac{G_F\,M_{Z}^2}{\sqrt2}
\left[ \left(v_n^q\right)^2 + \left(a_n^q\right)^2 \right]
 \delta\! \left(\hat s - M_{Z'}^2\right) \,.
\end{equation}
The $K$-factor represents the enhancement from higher order QCD processes,
estimated to be 
$K = 1 + {\alpha_s(M_{Z'}^2)\over2\pi} \textstyle {4\over3} \left( 1 +
{4\over3}\pi^2 \right) \simeq 1.3$ \cite{collider}.
When the mixing is ignored, 
$
\left(v_n^q\right)^2 + \left(a_n^q\right)^2 = (0.62)^2 \lambda
$
and the cross section is independent of the parameter $\gamma$ 
as long as  $\epsilon_V^2 + \epsilon_A^2 = 1$.

Note that all the current and previous dijet-mass searches at the Tevatron
are limited to $M_{jj}> 200$ GeV, which are not applicable to the present
$Z'$ with $M_{Z'} \approx 145$ GeV.
The relevant dijet data were from the UA2 Collaboration with collision 
energy at $\sqrt{s} =630$ GeV.  
The UA2 Collaboration\cite{ua2} has detected the $W+Z$ signal in the dijet-mass
region $48<m(jj)<138$~GeV and has placed upper bounds on $\sigma B(Z'\to jj)$
over the range $80<m(jj)<320$~GeV. 
The analysis against the UA2 data was shown in Fig.~1 of Ref.~\cite{bcl}.
We do not repeat the exercise here, but just use the result there. From 
Fig.~1 of Ref.~\cite{bcl} 
the allowed values are $\lambda \alt 1$ for
$M_{Z'} = 100 - 180$ GeV, given the uncertainty in the $K$-factor in 
the theoretical cross section calculation
and the difficulty in obtaining an experimental bound by
subtraction of a smooth background.
We shall consider $\lambda\alt 1$  in the following.

\begin{figure}[t!]
\includegraphics[width=3.5in]{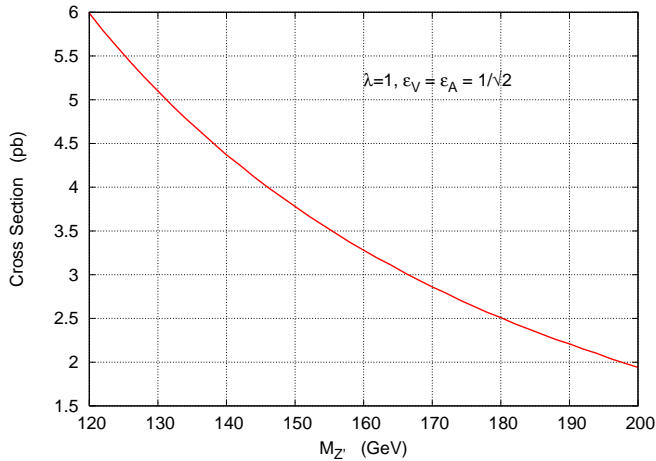}
\medskip
\caption{\small \label{fig1}
Production cross sections for $WZ'$ versus $M_{Z'}$ at the Tevatron 
($\sqrt{s} = 1.96$ TeV).  The parameters chosen are 
$\epsilon_V = \epsilon_A = 1/\sqrt{2}$ and $\lambda = 1$.
The SM $WZ$ production cross section is about 2.9 pb.
}
\end{figure}

{\it Associated Production.}--
The associated production of $Z'$ with a $W$ boson goes through the
$t$- and $u$-channel exchange of quarks while the $s$-channel boson
exchange is highly suppressed because of the negligible 
mixing angle between the SM $Z$ boson and the $Z'$.
Consequently, we expect similar or even larger cross sections for
$M_{Z'} \sim M_Z$ than the SM $WZ$ production in which there is a
delicate gauge cancellation among the $t$-, $u$-, and $s$-channel
diagrams.  The cross sections at the Tevatron energy $\sqrt s =
1.96$~TeV are shown for $\lambda=1$ in Fig.~\ref{fig1}.  We have
included a $K$-factor of $K=1.3$ to approximate next-to-leading order
QCD contributions \cite{ohnemus}.  We can see that at 
$M_{Z'}=140-150$ GeV the cross section is right at the order of 4 pb, which
is required to explain the excess in the CDF $Wjj$ anomaly
\cite{cdf-wjj}.  The choice of vector and axial-vector couplings are
\begin{equation}
\epsilon_V = \epsilon_A = \frac{1}{\sqrt{2}} \;,
\end{equation}
which are the same for up- and down-type quarks.  

The total width of the $Z'$ for $M_{Z'} = 145$ GeV is a mere 3 GeV. This is
not in contrast with the width observed \cite{cdf-wjj} because
the $M_{jj}$ distribution is dominated by the resolution.  We are not going
to fit our model to the $M_{jj}$ distribution (Fig.~1 of Ref.~\cite{cdf-wjj}),
because it can always be done by adjusting the bin resolution and the 
peak normalization. 

Since we have assumed here democratic couplings to all quarks,
$Z'$ can decay into $b\bar b$ with a branching ratio 
$B(Z' \to b\bar b) =0.2$.  Therefore, $WZ'$ production can give
rise to a $\ell \nu b\bar b$ final state.  Both CDF and D\O\  have
dedicated searches for $\ell \nu b \bar b$ final state for the Higgs boson
\cite{cdf-wh}. The preliminary $5.7$ fb$^{-1}$ result of CDF \cite{cdf-wh}
put 95\% C.L. upper limits on 
the ratio $\sigma(W X)\times B(X \to b\bar b)/ 
\sigma( W H_{\rm SM})\times B(H_{\rm SM} \to b\bar b)$.
For the particle with mass equal to 140, 145, and 150 GeV, the 
limits on the ratio are $15.8$, $25.3$, and $44.3$, respectively.
The $WZ'$ cross section here is about 4 pb for $M_{Z'} = 145$ GeV,
and with a branching ratio
$B(Z' \to b\bar b) =0.2$ the $\sigma(WZ') \times B(Z' \to b \bar b) \approx 
0.8$ pb. 
On the other hand, the SM $\sigma(WH) \times B(H\to b\bar b) \approx 21$ fb 
for $m_H = 145$ GeV.  Thus, the largest allowed cross section 
for $\sigma(WZ') \times B(Z'\to b \bar b) \approx 0.53$ pb.
Given the uncertainty in all these estimations, 
the current upper limit on
$\sigma(WH) \times B(H\to b \bar b)$ from the Tevatron 
begins to constrain
the $Z'$ model. 
If we give up the simple assumption of democratic choice on 
$Z'$-$q$-$\bar{q}$ couplings to all generations,
we can easily lower the $Wb\bar{b}$ event rate.
Also, note that the amount to be reduced is mere.

\begin{figure}[t!]
\includegraphics[width=3.5in]{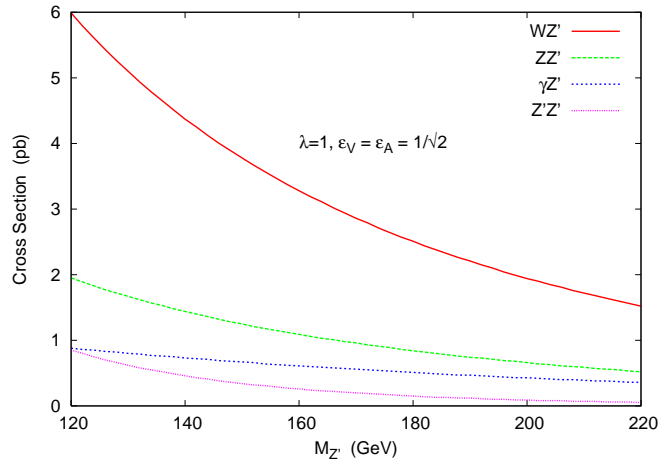}
\caption{\small \label{fig2}
Production cross sections for $WZ'$, $ZZ'$, $\gamma Z'$, and $Z' Z'$ 
versus $M_{Z'}$ at the Tevatron ($\sqrt{s} = 1.96$ TeV).  The parameters 
chosen are  $\epsilon_V = \epsilon_A = 1/\sqrt{2}$ and $\lambda = 1$.
We have imposed the acceptance $p_T(\gamma)>50$ GeV and $|\eta(\gamma)|<1.1$ 
on the final state photon.
}
\end{figure}

{\it Implications at the Tevatron.}--
As shown in Ref.~\cite{bcl} other associated production channels,
$\gamma Z',\; Z Z'$, and $Z' Z'$ are possibly observable, provided
that the current excess is due to $WZ'$ production.
We show the production cross sections for these channels in Fig.~\ref{fig2}.
We have imposed the following acceptance
on the final state photon \cite{photon:cut}:
\begin{equation}
\label{photon}
p_T(\gamma)>50 \;{\rm GeV},\;\;\; |\eta(\gamma)|<1.1 .
\end{equation}

The irreducible backgrounds to the $(\gamma,W,Z)Z'$ signals with $Z'\to  jj$
arise from the $(\gamma,W,Z) jj $ final states.
An analysis with $jj = b \bar b$ had been performed in Ref.~\cite{bcl}.
The $WZ'$ signal has the advantage that its significance in the presence
of the corresponding background is better than the other channels.
We expect that a similar advantage is enjoyed by the $WZ' \to Wjj$ mode.

It was mentioned in Refs.~\cite{cdf-wjj,cdf-Z} and in Ref.~\cite{cdf-gamma}
that no significant excess is observed in $Zjj$ and $\gamma jj$ channel,
respectively. We shall show that with current systematic uncertainties
of level 10\% \cite{cdf-wjj,cdf-Z,cdf-gamma,thesis} and a similar set of
jet cuts, no significant excess can be observed in both channels.
% in accord with Refs.~\cite{cdf-wjj,cdf-Z,cdf-gamma}.
With the jet cuts
$E_{Tj} > 30 \;{\rm GeV}, \; |\eta_j| < 2.4, \; p_{Tjj} > 40\;{\rm GeV},
\; 120\;{\rm GeV} < M_{jj} < 160\; {\rm GeV}$ and leptonic cuts
$p_{T\ell} > 20\;{\rm GeV},\; |\eta_\ell| < 2.8$, the 
$\sigma_{\rm signal}: \sigma_{\rm bkgd} = 26 \,{\rm fb}: 171\,{\rm fb}$ 
for the $Zjj$ channel. It would give a significance of
$S/(\sqrt{B} \oplus 0.1 B) \approx 1.4\sigma$ for ${\cal L}=4.3\,
{\rm fb}^{-1}$, where the factor $0.1$ is the systematic uncertainties. 
With the same set of jet cuts and photon cuts in Eq.~(\ref{photon}) 
to the $\gamma jj$ channel,
we obtain $\sigma_{\rm signal}: \sigma_{\rm bkgd} = 0.5 \,{\rm pb}: 9.9\,{\rm pb}$,
which gives a significance of 
$S/(\sqrt{B} \oplus 0.1 B) \approx 0.5\sigma$.
Therefore, we cannot observe any significant excess in both channels,
in accord with the claims in Refs.~\cite{cdf-wjj,cdf-gamma}.

Nevertheless, if we tighten the jet cuts the backgrounds will suffer 
more than the signals. With $E_{Tj} > 50\,{\rm GeV}$ and 
${\cal L}=10\, {\rm fb}^{-1}$, the significance can improve to 
$2.3\sigma$ and $1\sigma$ for $Zjj$ and $\gamma jj$ channel, respectively.
If the systematic uncertainties can be reduced to an ideal level of $2\%-3$\%
the significance can be further improved to $5\sigma$ and $4\sigma$,
respectively.  Details will be presented later \cite{future}.

{\it Conclusions.}--
We have shown that a baryonic $Z'$ boson can explain the
excess in the invariant-mass window $120 - 160$ GeV in the 
dijet system of $Wjj$ production.  Such a $Z'$ boson with depleted 
leptonic couplings is not subject to the current dilepton limits on extra 
gauge bosons.  Yet, the strongest constraint comes from the dijet 
search of the UA2 data, from which the size of coupling, proportional
to $\sqrt{\lambda}$, 
is constrained to be $\lambda \alt 1$.  With $\lambda =1$ we are
able to explain the required cross section of 4 pb in the excess
window. 
We have also shown that it is hard to see the excess in
both $\gamma Z' \to \gamma jj$ and $Z Z' \to \ell^+ \ell^- jj$ channels
under the current systematic uncertainties and jet cuts.
However, with tightened jet cuts and improved systematic uncertainties
it could be promising to test the excess in these two channels.

Other comments and possibilities are given as follows:
\begin{itemize}
\item[(i)]
A more dedicated dijet analysis at the energy range $100 - 200$ GeV 
at the Tevatron could be another important test for this baryonic
$Z'$ model.  

%\item
%In Ref.~\cite{cdf-wjj}, it was said that the $Z+$jets sample did not
%show significant deviations from the SM.  It can be understood from the
%fact that the $WZ'$ signal in the presence of $Wjj$ 
%background enjoys a higher significance than the $ZZ'$ signal in the
%presence of $Zjj$ background, and similarly for $\gamma Z'$ signal.
%We have shown in this work that with more dedicated searches the 
%$\gamma jj$ could be promising.

\item[(ii)]
A baryonic $W'$ boson is equally possible to explain the anomaly, although
the size of coupling would be different.
In addition, this $W'$ has an additional advantage since
the constraint from the current upper limit on
$\sigma(WH) \times B(H\to b \bar b)$ at the Tevatron
does not apply to $m_{W'} < m_t+m_b$.
However, it would also be subject to the 
dijet constraint of UA2.  Similarly, it would predict excess in 
$\gamma W'$ and $Z W'$ production.
\item[(iii)]
Another interesting kinematics to look at is the angular distribution
of the scattering angle $\theta_{sc}$. Similar to the SM $WZ$ production, 
the scattering angle of the $Z'$ would also be peaked at 
$|\cos \theta_{sc}| = 1$.  On the other hand, a Higgs-like boson would have
a flat distribution in $\cos\theta_{sc}$. 
\item[(iv)]
The prospects for detecting the $Z'$ would be best in the $Z' \to b\bar b$
final state, with $b$-tagging by vertex detector or semileptonic decays to
reject backgrounds from light quarks and gluons in the $(\gamma,W,Z)jj$ final
state.
\item[(v)]
Dedicated searches on $\ell \nu b \bar b$ or $\ell \bar \ell b \bar b$ also
provide useful tests for the model.  As long as the new $Z'$ does not
have suppressed couplings to $b \bar b$, such searches will begin
to probe the useful range of the parameters. 
\end{itemize}

We thank Vernon Barger and Paul Langacker for collaboration of the $Z'$
project 15 years ago that led to the present Letter, and thank to
Shin-Shan Yu (Eiko) for information on 
current $\gamma jj$ search.
This research was supported in parts by the NSC under Grant
No. 99-2112-M-007-005-MY3 and by WCU program through the NRF funded by 
the MEST (R31-2008-000-10057-0).

{\it Note added.}--A few papers \cite{new} appeared one 
day after the appearance of Ref.~\cite{cdf-wjj}. We share similar ideas, 
though in a different framework. Also, there were some related 
works \cite{before} before that.

%%%%%%%%%%%%%%%%%%%%%%%%%%%%%%%%%%%%%%%%%%%%%%%%%%%%%%%%%%

\end{document}